# Polarization-cloaked spatiotemporal topological knots

Yaning Zhou[1,2†], Nianjia Zhang[3†], Zhao Zhang[1,2], Chunhao Liang[1,2], Qian Cao[3], Houan Teng[4], Qiwen Zhan[4,5*], Sergey A. Ponomarenko[6,7*], Xu Guo[8,*], Yangjian Cai[1,2*], Xin Liu[1,2*]

[1]Shandong Provincial Key Laboratory of Light Field Manipulation Physics and Applications & School of Physics and Electronics, Shandong Normal University, Jinan 250014, China

[2]Collaborative Innovation Center of Light Manipulations and Applications, Shandong Normal University, Jinan 250358, China

[3]School of Optical-Electrical and Computer Engineering, University of Shanghai for Science and Technology, Shanghai 200093, China

[4]Zhejiang Key Laboratory of 3D Micro/Nano Fabrication and Characterization, Department of Electronic and Information Engineering, School of Engineering, Westlake University, Hangzhou 310030, China

[5]Westlake-Kromoson Joint Laboratory of Light-Field Intelligence, School of Engineering, Westlake University, Hangzhou 310030, China

[6]Department of Physics and Atmospheric Science, Dalhousie University, Halifax B3H 4R2, Canada

[7]Department of Electrical and Computer Engineering, Dalhousie University, Halifax B3J 2X4, Canada

[8]Geotechnical and Structural Engineering Research Center, Shandong University, Jinan 250061, China

[†]These authors contributed equally to this work.

[*]Corresponding authors: *zhanqiwen@westlake.edu.cn*;*serpo@dal.ca*; guoxu@sdu.edu.cn; *yangjiancai@sdnu.edu.cn*; *xinliu@sdnu.edu.cn*

**Spatiotemporally localized topological knots encode global connectivity within a moving space-time optical wavepacket, circumventing the requirement for spatial-only eigenmodes that extend throughout three-dimensional space-filling. In conventional modalities, this connectivity is identified directly from a scalar complex field. However, knotting optical knots with the polarization state of light proves difficult to realize for ultrafast localized wavepackets. In this work, we combine oppositely wound toroidal vortex modes with orthogonal circular polarization states on the higher-order Poincaré sphere to construct vectorial electromagnetic spatiotemporal wavepackets, in which the topological connectivity is encoded in the relative phase between the polarization components. Stokes analysis shows that $S_0$ retains a common toroidal envelope, whereas the phase sensitive Stokes parameters $S_1$ and $S_2$ reveal winding-dependent linked and knotted trajectories; numerical**

**simulations verify their closure, and time-sliced off-axis interferometry experimentally reconstructs the corresponding three-dimensional $S_2$ textures. This separation of topological connectivity from the scalar intensity envelope extends the control of spatiotemporal topology to the polarization degree of freedom and provides a versatile framework for controlling how topological information is accessed in multidimensional structured light.**

## Introduction

Optical links and knots are topologically structured configurations of closed curves embedded in light fields. These locus curves typically correspond to phase singularity lines, where the field amplitude vanishes and the phase becomes undefined [1-3]. Conventional scalar optical knots are generally implemented through paraxial propagation of transverse eigenmodes, yielding three-dimensional (3D) structures that extend along the propagation direction [1,2,4-6]. The introduction of framed knots has further expanded the accessible configurations [7]. At the same time, the influence of aberrations, turbulence and partial coherence on their structural integrity has been studied [8-10]. Despite these advances, topologies reported in such scalar approaches are identified in the spatial trajectories of field singularities. Extending optical topology into the spatiotemporal domain introduces local time as an additional coordinate, opening new possibilities for shaping and confining topological structures within propagating wavepackets [11]. Spatiotemporal optical vortices carry phase singularities in the coupled 2D space-time plane [12,13], whereas 3D localized space-time wavepackets confine their structured optical fields within a finite volume in $(x, y, \tau)$ [14,15]. Toroidal light vortices (TLVs) form a closed vortex geometry characterized by independently controllable poloidal and toroidal winding numbers [16-18]. These advances have paved the way for the spatiotemporal localization of scalar links, knots and hopfion textures [19,20]. Yet, the topology of localized optical links and knots is still identified directly from iso-intensity surfaces or local intensity maxima [19]. Therefore, in this sense, spatiotemporal localization transforms where optical topology resides, but not how it is encoded and revealed. This raises a fundamental question: Can a localized link-knot texture remain cloaked in the total intensity, with its topology encoded in—and revealed exclusively through— the *polarization* structure of the space-time wavepacket?

Polarization offers a fundamentally different route to encoding and revealing optical topology by exploiting the vectorial structure of light. The higher-order Poincaré sphere (HOPS) describes coherent superpositions of orthogonally polarized structured modes [21-23] and the concept has been extended to spatiospectral [24] and spatiotemporal vector fields [25,26]. In the circular polarization basis, modal intensity ratio determines polarization ellipticity, while relative phase governs polarization azimuth [21,22,27]. Accordingly, the Stokes parameters $S_0$ and $S_3$ quantify modal intensities, whereas $S_1$ and $S_2$ characterize their relative phase via orthogonal coupling. These principles underpin diverse forms of polarization-based topology of light, including Möbius strips [28-30] and knotted C-lines [27,31], as well as skyrmions and hopfions defined by global mappings of normalized electromagnetic or Stokes vectors [31-34]. Their extension to spatiotemporal skyrmions [35,36] and space-time hopfion crystals [37] further expands the topological landscape of structured light. Meanwhile, Stokes-selective holography has demonstrated the recovery of image information inaccessible through intensity measurements [38], complementing boarder advances in higher-dimensional structured light and optical topological textures [39-43]. Despite this progress, a conceptual distinction remains between topology carried by polarization itself and information encoded in polarization-resolved measurements. Consequently, it remains unknown whether distinct spatiotemporal link-knot textures can share the same total intensity envelope. More specifically, can their connectivity be encoded within the polarization structure of a non-uniformly polarized electromagnetic spatiotemporal wavepacket (ESTWP)? The central question is whether the connectivity of scalar toroidal vortex fields can be written into a HOPS field—remaining cloaked in $S_0$, while being readable through phase-sensitive Stokes parameters.

Here, we combine the topology of scalar toroidal vortices with the vectorial degrees of freedom of the HOPS to encode spatiotemporal topological textures in the polarization structure of ESTWPs [16,19,26]. Specifically, two conjugate TLV modes with opposite poloidal and toroidal windings are assigned to orthogonal circular polarization channels. Their modal intensities determine $S_0$ and $S_3$, preserving the shared toroidal envelope, whereas their relative phase gives rise to interference structures in $S_1$ and $S_2$, where linked and knotted trajectories emerged. Tuning the toroidal winding parameter causes these trajectories to close either as two mutually linked loops or as a single knot with a prescribed winding number. We map their closure onto an unfolded braid representation and reconstruct the 3D $S_2(x, y, \tau)$ distributions through time-sliced off-axis interferometry [19]. The topology is therefore veiled rather than absent: while the total intensity displays only the common toroidal envelope, the relevant Stokes parameters unveil the nontrivial connectivity. Polarization thus dictates both the local vector state and the observable through which the spatiotemporal topology is extracted.

## Results

### Winding controlled Stokes links and knots

Orthogonally polarized modes provide a complete basis for constructing structured optical wavepackets. Each optical polarization state can be represented by a point on the HOPS, where the north and south poles correspond to right- and left-hand circularly polarized states, respectively. Within this framework, we consider a classically entangled HOPS state of an ESTWP, expressed as follows:

$$\boldsymbol{E}(\boldsymbol{r},\tau)=\cos\left(\frac{\Theta}{2}\right)e^{-i\Phi/2}\Psi_{-}(\boldsymbol{r},\tau)\boldsymbol{e}_{-}+\sin\left(\frac{\Theta}{2}\right)e^{i\Phi/2}\Psi_{+}(\boldsymbol{r},\tau)\boldsymbol{e}_{+}, \tag{1}$$

where $\boldsymbol{e}_{\pm}=\left(\boldsymbol{e}_x\pm i\boldsymbol{e}_y\right)/\sqrt{2}$. The angles $\Theta$ and $\Phi$ determine the relative modal weights and phase on the HOPS, respectively. Here, $\boldsymbol{r}=(x,y)$, and $\tau=t-z/v_g$ denotes the local time in a reference frame moving with the group velocity, $\tau$ being expressed in the normalized units used for the space-time mapping. To encode the spatiotemporal winding in the polarization structure, we choose the two modes to be toroidal vortices with opposite poloidal and toroidal phase windings,

$$\Psi_{\pm}(\mathbf{r},\tau)\propto A(r_{\perp})\exp(\mp il\phi_{\perp})\exp(\mp im\phi_{\parallel}), \tag{2}$$

where $A(r_{\perp})=r_{\perp}^{|l|}\exp(-r_{\perp}^2/w_0^2)$ . $r_{\parallel}=\sqrt{x^2+y^2}$ and $\phi_{\parallel}=\tan^{-1}(y/x)$ are the polar coordinates in the transverse plane, while $r_{\perp}=\sqrt{(r_{\parallel}-r_0)^2+\tau^2}$ and $\phi_{\perp}=\tan^{-1}[\tau/(r_{\parallel}-r_0)]$ are defined in the poloidal space-time plane. The topological charges $l$ and $m$ govern the phase winding in the poloidal and toroidal directions, respectively. The Stokes parameters are defined as $S_0=|E_+|^2+|E_-|^2$, $S_3=|E_+|^2-|E_-|^2$ and $S_1+iS_2=2E_+^*E_-$. Substituting Eqs. (1) and (2) into these relations yields the expressions

$$S_0\propto A^2(r_{\perp}), \tag{3}$$

$$S_1\propto A^2(r_{\perp})\sin\Theta\cos(2l\phi_{\perp}+2m\phi_{\parallel}-\Phi), \tag{4}$$

$$S_2\propto A^2(r_{\perp})\sin\Theta\sin(2l\phi_{\perp}+2m\phi_{\parallel}-\Phi), \tag{5}$$

$$S_3\propto -A^2(r_{\perp})\cos\Theta. \tag{6}$$

Equations (3)-(6) provide the analytical basis for the Stokes structures in Fig. 1. The common toroidal envelope is retained by $S_0$ (Fig. 1b) and $S_3$ (Fig. 1e), whereas the doubled relative phase appears in the phase sensitive observables $S_1$ (Fig. 1c) and $S_2$ (Fig. 1d). The factor $\sin\Theta$ controls their amplitude, whereas $\Phi$ shifts the phase origin of the resulting structures without changing the winding parameters. These different dependencies lead to the contrasting Stokes geometries shown in Fig. 1. The isosurfaces of $S_0$ and $S_3$ retain the toroidal envelope of the wavepacket, whereas $S_1$ and $S_2$ form displaced, intertwined tubular structures within the same $(x,y,\tau)$ region. Because the two observables select the cosine and sine quadratures of the same relative phase, their structures differ by a $\pi/2$ phase offset rather than by a different winding. The surface hue encodes the doubled polarization azimuth $2\psi = \arg(S_1 + iS_2)$, and therefore traces the local orientation of the polarization ellipse modulo $\pi$.

The connectivity of the intertwined structures becomes clearer when the closed torus is unfolded into a periodic cylinder (Fig. 1f). The toroidal angle $\phi_{\parallel}$ is mapped onto the cylinder axis, while the poloidal angle $\phi_{\perp}$ describes rotation around its cross section. The center trajectories of the Stokes tubes are thereby represented by interlaced strands. When the cylinder is closed by joining its two ends, phase continuity determines how their endpoints are joined: joining endpoints on the same branch produces separate linked loops, whereas joining the neighboring phase shifted branches produces a single closed knot. In the present framework, polarization cloaking refers to the separation between the scalar envelope and the relative phase topology. The total intensity retains the common toroidal envelope but does not resolve the connectivity carried by the relative phase; this information is accessed through the polarization interference observables $S_1$ and $S_2$. We next vary the toroidal winding parameter to examine how it changes the braid closure and the resulting spatiotemporal links and knots.

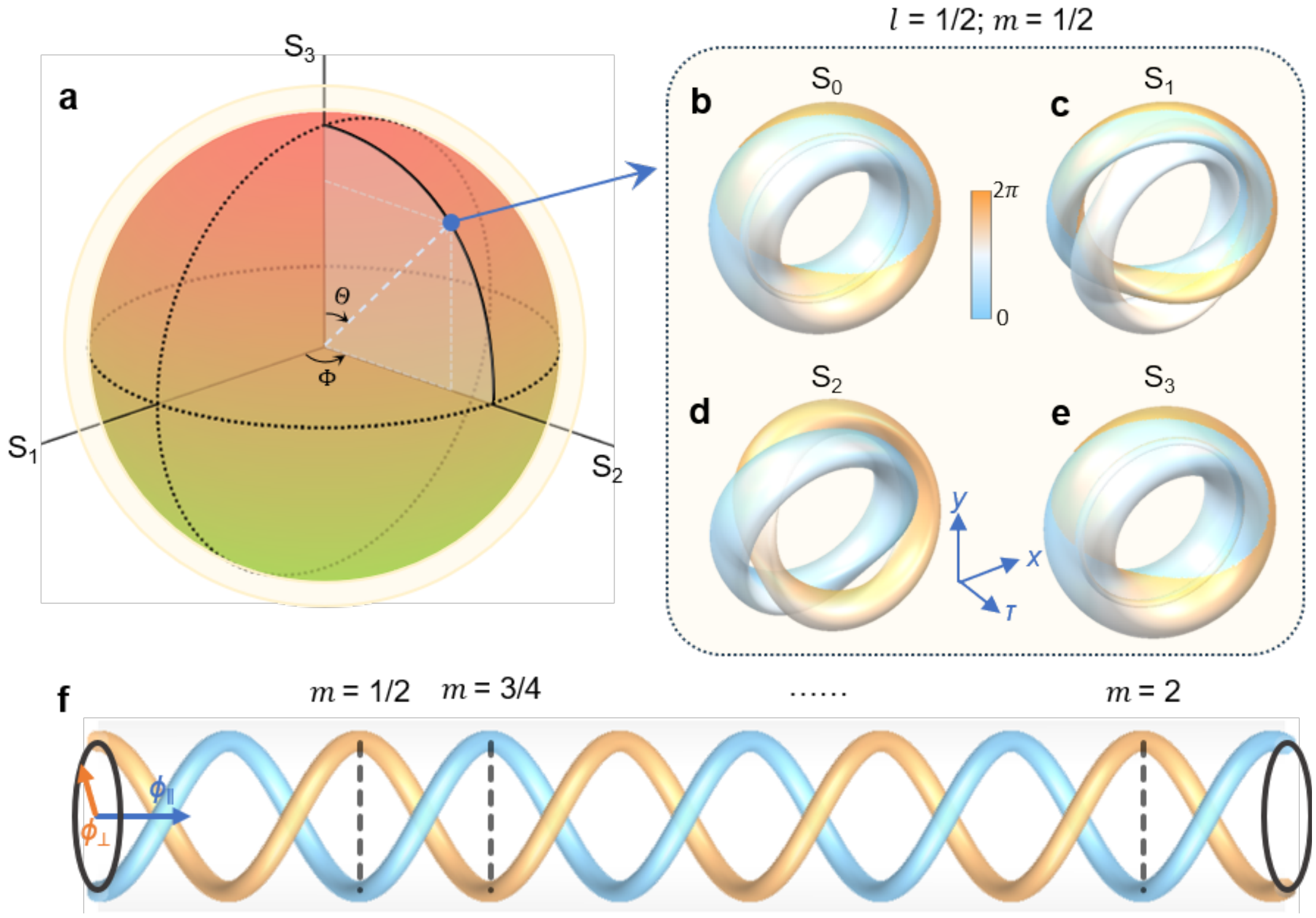

**Fig. 1. Illustration of polarization-cloaked spatiotemporal topology on the HOPS. a**, HOPS representation of two TLV modes with orthogonal circular polarizations, where $\Theta$ and $\Phi$ are the polar and azimuthal angles of the spherical coordinates. **b-e**, 3D isosurfaces of $S_0$, $S_1$, $S_2$, and $S_3$. The toroidal envelopes are manifested in $S_0$ and $S_3$, whereas $S_1$ and $S_2$ reveals intertwined structure. The surface color represents the doubled polarization azimuth $2\psi = \arg(S_1 + iS_2)$. **f**, Cylindrical braid representation obtained by unfolding the torus along the toroidal angle $\phi_{\parallel}$.

To examine how the phase winding controls the connectivity of the polarization-cloaked structures, we use $S_2$ as the phase sensitive observable. We fix the HOPS state at $\Theta = \pi/4$ and $\Phi = \pi/2$ and set the poloidal parameter to $l = 1/2$, while varying only the toroidal parameter $m$. This keeps the common toroidal envelope unchanged while tuning the phase accumulation along the toroidal direction. We first consider the linked configurations obtained for $m = 1/2$ and $3/2$ (Fig. 2a,b). In both cases, the $|S_2|$ isosurfaces comprise two independent, well-defined closed tubes localized within the 3D $(x, y, \tau)$ domain. The two components wind around one another without merging, and the larger value of $m$ gives a greater number of turns around the torus. The corresponding trajectory winding orders are 2 and 6, respectively. For $m = 3/4$ and $5/4$, the trajectories close differently (Fig. 2c,d). Here, the two phase-shifted branches no longer close independently but join into a single continuous tube, forming a single component knot. Increasing $m$ increases the number of turns around the toroidal envelope, yielding winding orders of 3 and 5. The second viewing direction makes the single component closure more apparent and distinguishes the knot from a projected overlap of two separate loops.

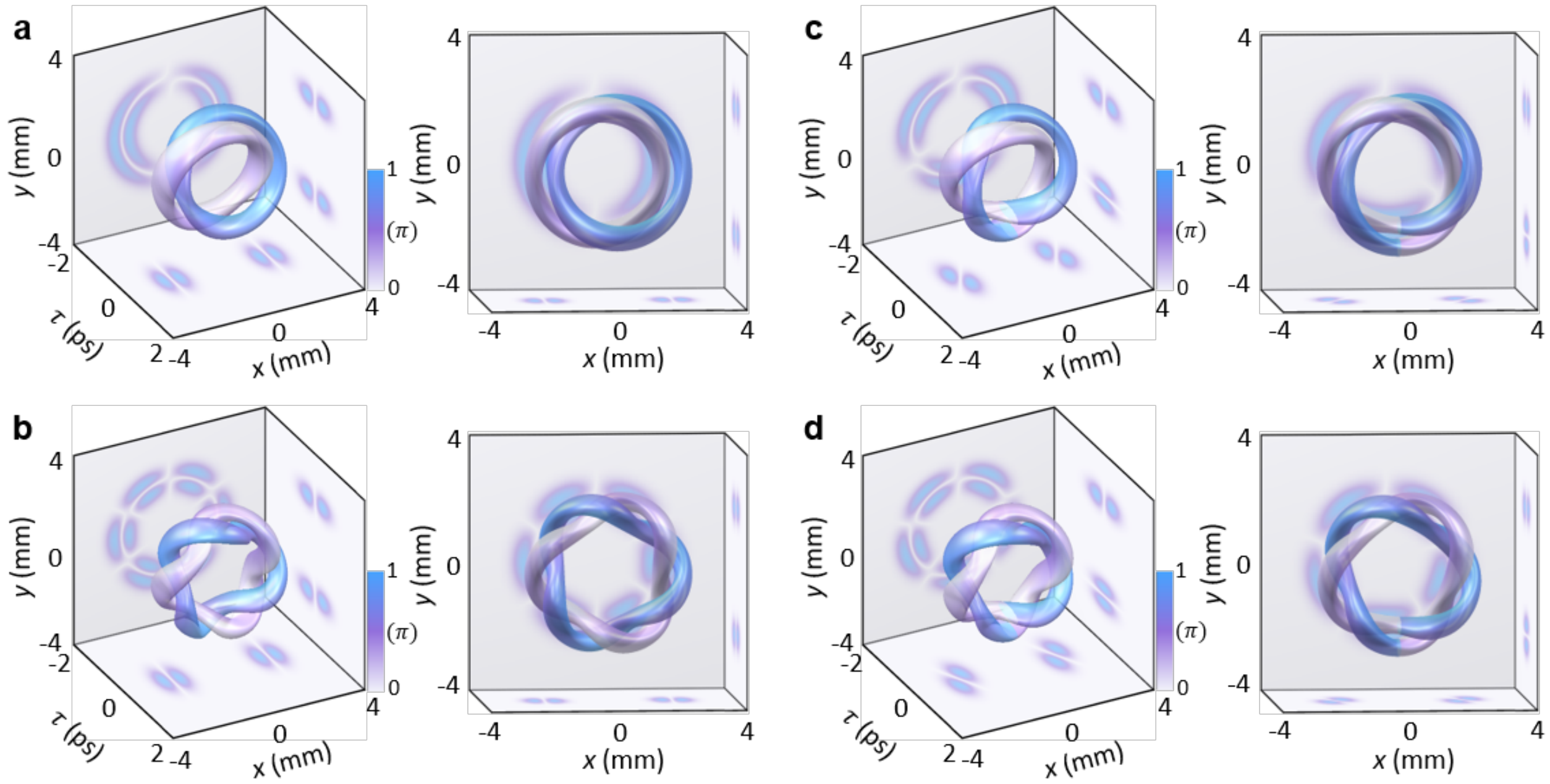

**Fig. 2. Weaving optical Stokes links and knots.** 3D $|S_2|$ isosurfaces calculated for $l = 1/2$, $\Theta = \pi/4$, and $\Phi = \pi/2$. The isosurface threshold is set to 40% of its peak value. **a**, $m = 1/2$, link with winding number 2. **b**, $m = 3/2$, link with winding number 6. **c**, $m = 3/4$, knot with winding number 3. **d**, $m = 5/4$, knot with winding number 5. The left panels include the $x = 0$, $y = 0$, and $\tau = 0$ sections, while the right panels show the structures from alternative viewing directions. The surface color indicates twice the local polarization azimuth angle $2\psi$.

**Experimental generation of TLVs and topology reconstruction**

The change in connectivity follows from the relative phase $\chi = 2l\phi_\perp + 2m\phi_\parallel - \Phi$ encoded in the phase sensitive $S_2$ distribution. For $l = 1/2$, the effective poloidal winding is unity, whereas $2m$ determines the phase accumulation around the toroidal direction. For an integer $2m$, each of the two branches to itself after one toroidal circuit, producing two separately closed but mutually intertwined components. For a half-integer $2m$, the two branches are exchanged at the periodic boundary and join into a single closed trajectory. Thus, varying the toroidal phase winding changes the link-knot connectivity while leaving the HOPS weights and the common spatiotemporal envelope unchanged. The surface color records the local polarization azimuth and shows its variation along each tube, whereas the link-knot distinction is determined by the closure of the phase branches rather than by the color distribution. These results show that varying the winding parameter produces a family of polarization-cloaked Stokes links and knots with different connectivities and winding numbers. We next investigate whether these theoretically prescribed structures can be reconstructed from experimentally retrieved TLV fields. The polarization-sensitive topology depends on the relative phase between the constituent toroidal modes, making time resolved complex field measurements necessary for its experimental reconstruction. We implemented this measurement by combining HOPS-weighted TLV pulse shaping with time-sliced off-axis interferometry to separately retrieve the complex fields of the right- and left-circularly polarized components, as illustrated in Fig. 3. A pulsed beam centered at 1030 nm, with a spectral bandwidth of approximately 15 nm, was divided by BS1 into an object arm for TLV synthesis and a reference arm for spatiotemporal field retrieval.

In the object arm, a two-dimensional pulse shaper comprising a diffraction grating, cylindrical lens CL1, and SLM1 first generated a poloidal spatiotemporal optical vortex. SLM1 encoded the phase associated with the poloidal winding, while the afocal cylindrical expander formed by CL2 and CL3 stretched the vortex along one transverse direction with a magnification of 1:7, producing an extended vortex tube. SLM2 subsequently applied the conformal transformation to bend the tube into a closed toroidal geometry. The toroidal phase was then imposed by SLM3, which also compensated the residual wavefront curvature before the wavepacket was delivered to the detection plane. For the two successive measurements, the pulse-shaping holograms were designed to generate the complex mode profiles prescribed by the two HOPS components, including a common toroidal amplitude envelope, opposite poloidal and toroidal phase windings, and the specified modal weights and phase factors. These weights and phase factors were determined by the HOPS angles Θ and Φ, respectively. A quarter-wave plate (QWP) placed after SLM3 converted the linearly polarized output into the right- or left-circularly polarized state corresponding to the encoded mode. The reference arm provides the time resolved sampling required to retrieve the 3D field. A grating pair generated a Fourier-transform-limited reference pulse, and a right-angle prism mounted on the delay line controlled its temporal overlap with the object pulse. The two pulses were recombined at BS4 with a small transverse angular offset and recorded by the camera. The circularly polarized object field interfered with the matching circular polarization component of the linearly polarized reference pulse, while the orthogonal reference component contributed only to the intensity background. The reference pulse thus served as a temporal gate, with each delay position sampling a temporal slice of the object wavepacket. The angular offset introduced a spatial carrier into the interference fringes, allowing the complex field of each slice to be retrieved by Fourier filtering.

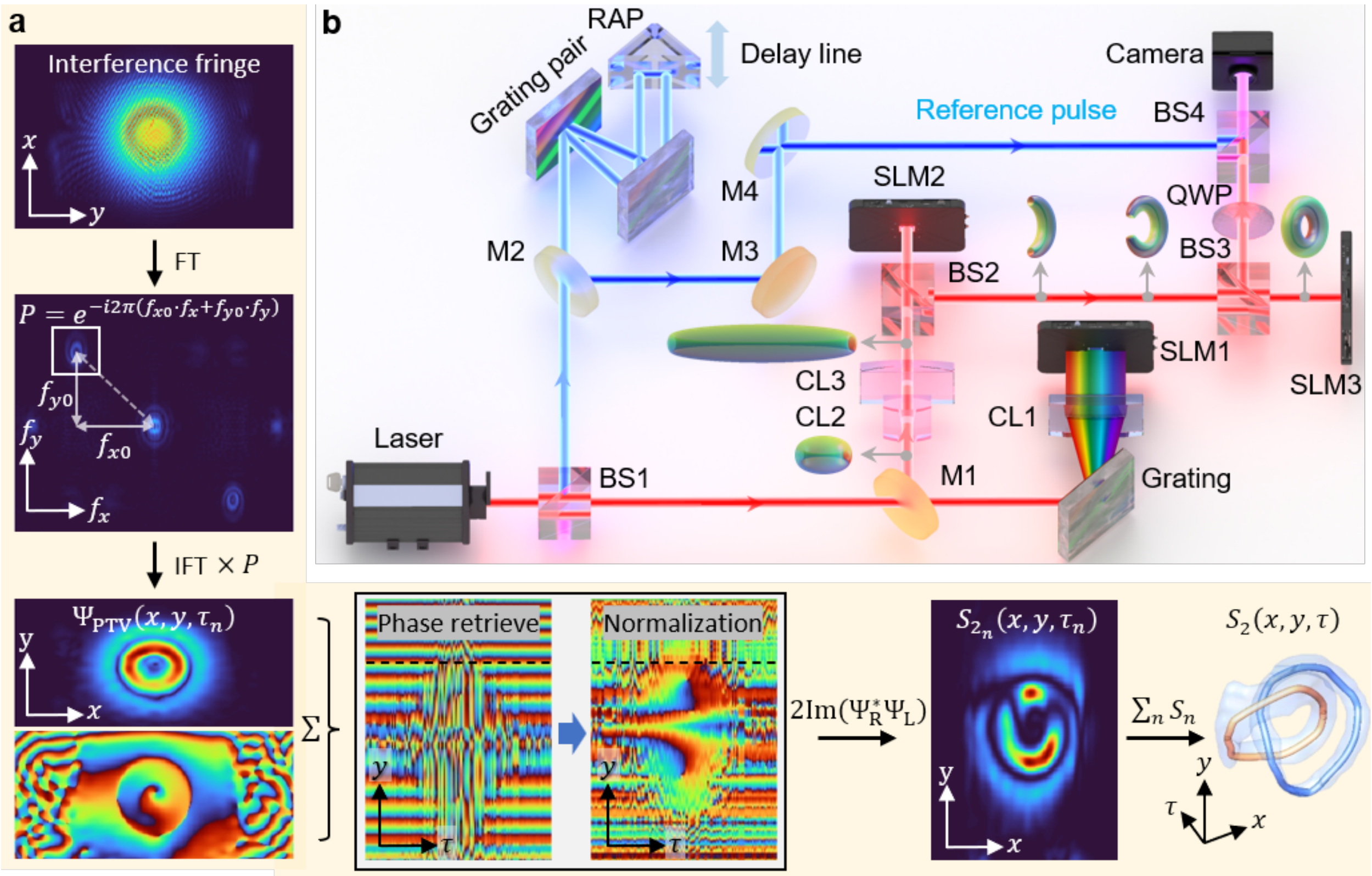

**Fig. 3 | Experimental setup for HOPS-encoded TLV generation and optical Stokes topology reconstruction. a**, Time-sliced off-axis interferometric reconstruction. Each interferogram recorded in the spatial $(x, y)$ plane is Fourier transformed into the $(f_x, f_y)$ plane, where a first-order sideband is selected. Carrier removal followed by inverse Fourier transformation recovers the complex field at each delay. Phase-referenced slices from 120 delays are assembled along τ to

separately reconstruct the 3D complex fields of the two HOPS-encoded circular polarization modes, from which $S_2(x, y, \tau)$ is numerically calculated. **b**, Experimental setup for the generation and characterization of HOPS-encoded circularly polarized TLV modes. A 1030 nm pulse is shaped by three SLMs, converted into circular polarization by a QWP, and interfered with a delayed Fourier-transform-limited reference pulse. BS, beam splitter; M, mirror; CL, cylindrical lens; SLM, spatial light modulator; RAP, right-angle prism; QWP, quarter-wave plate.

The reconstruction procedure is summarized in Fig. 3a. At each delay $\tau_n$, the camera recorded an interferogram in the spatial $(x, y)$ plane. After subtracting the independently measured, time integrated intensities of the object and reference pulses, the remaining interference signal was transformed into the spatial frequency $(f_x, f_y)$ plane. The off-axis carrier separated the conjugate first-order sidebands from the central component, with their positions determined by the fixed carrier frequencies $f_{x0}$ and $f_{y0}$. One first-order sideband was isolated, and its frequency offset was removed prior to inverse Fourier transformation. The inverse Fourier transform returned the selected sideband to the spatial $(x, y)$ plane, where its amplitude and phase provided the complex field sampled at the selected temporal delay. Any residual linear phase associated with imperfect carrier removal was corrected in this spatial domain field. The same Fourier-filtering procedure was applied independently to retrieve the complex fields of the two encoded circular polarization modes. For each encoded polarization mode, the retrieval was repeated over 120 temporal delays spanning the complete object pulse. Small variations in the fringe position between successive delays introduced random phase offsets among the reconstructed slices. To reduce phase fluctuations between successive temporal slices, each phase slice was normalized with respect to the same fixed spatial position. The phase referencing was performed separately for the two measurements, while their constant relative phase was specified in the subsequent numerical synthesis. This procedure reduced the delay dependent phase offsets and allowed the two-dimensional complex field slices to be assembled along $\tau$, yielding the reconstructed 3D complex field of each encoded polarization mode.

The independently reconstructed complex fields of the two HOPS-weighted TLV modes were then assigned to their respective circular polarization channels to construct the vectorial spatiotemporal field. For each temporal delay, the imaginary part of the cross term between the two reconstructed circular polarization components yielded the corresponding two-dimensional $S_2(x, y, \tau)$ distribution. Arranging the 120 reconstructed slices in chronological order produced the 3D $S_2(x, y, \tau)$ volume shown schematically at the end of Fig. 3a. This procedure uses the experimentally retrieved complex fields of the two HOPS-encoded polarization modes to numerically reconstruct the winding-dependent Stokes structure. Next, we present the linked and knotted configurations reconstructed through this approach.

The reconstructed $S_2(x,y,\tau)$ distributions recover the parameter dependent topological textures predicted in Fig. 2. Figure 4 presents the results for $m=1/2,\ 3/2,\ 3/4$ and $5/4$, with $l=1/2,\ \Theta=\pi/4$, and $\Phi=\pi/2$ kept unchanged. We plot the outer semi-transparent $|S_2|$ isosurface at 5% of the corresponding peak value to show the spatial and temporal extent of each reconstructed structure. The orange and blue solid lines trace the local maxima of the reconstructed $|S_2|$ distributions and identify the trajectories within the tubular isosurfaces. Additional viewing directions and orthogonal cross sections show the same reconstructed structures from different perspectives.

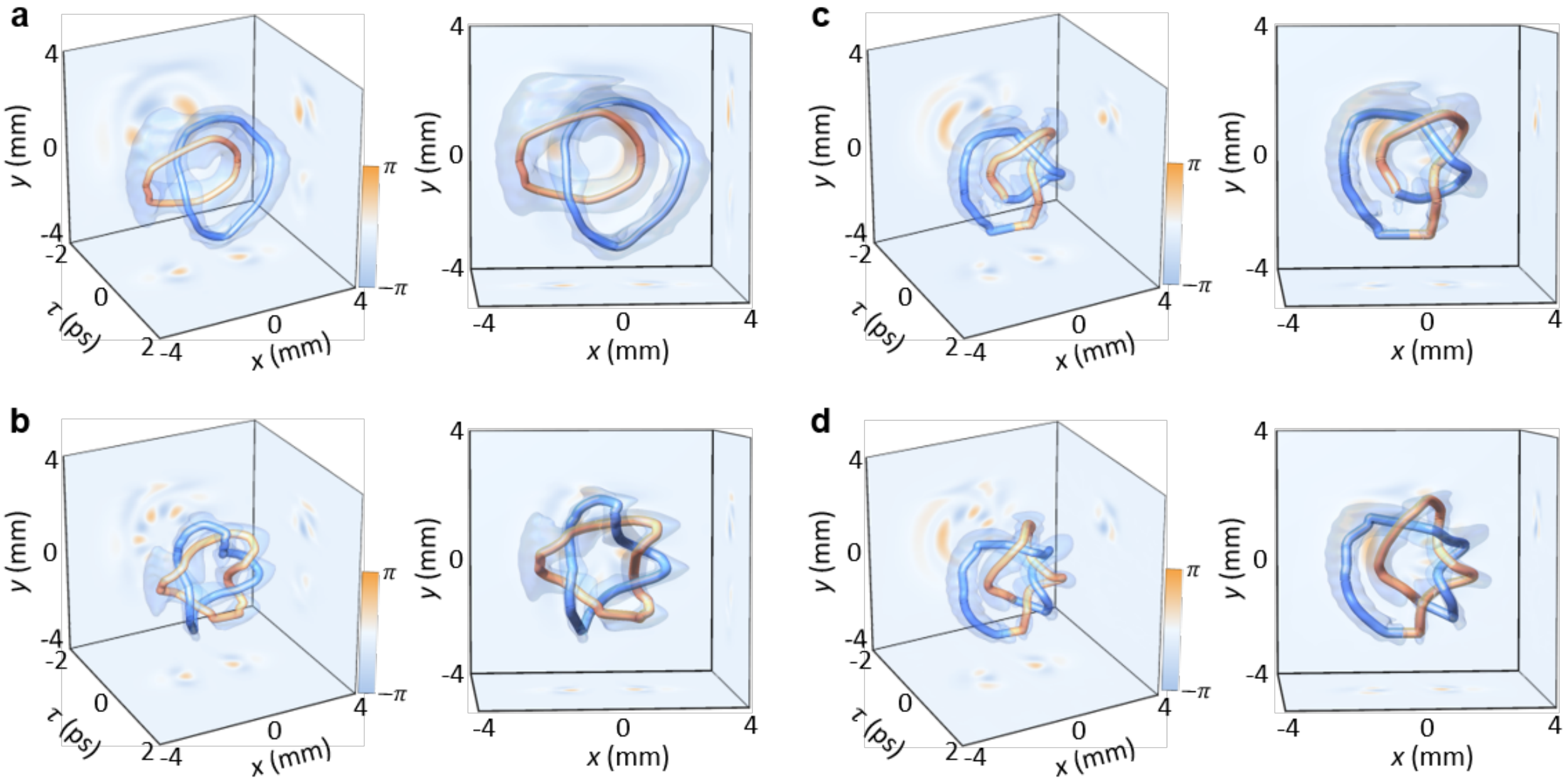

**Fig. 4. Experimental demonstration of polarization-cloaked links and knots.** Reconstructed $S_2(x,y,\tau)$ distributions for **a**, $m=1/2$; **b**, $m=3/2$; **c**, $m=3/4$; and **d**, $m=5/4$, with the remaining parameters identical to Fig. 2. The outer semitransparent $|S_2|$ isosurface is plotted at 5% of the peak value. Orange and blue curves trace the local maxima of $|S_2|$. The linked structures have winding numbers 2 and 6, whereas the knotted structures have winding numbers 3 and 5. Orthogonal sections and additional viewing directions further show the 3D connectivity of the reconstructed trajectories.

For $m=1/2$ and $m=3/2$, the reconstructed $|S_2|$ isosurfaces form two separate closed tubular components localized in the $(x,y,\tau)$ domain. The corresponding closed trajectories remain spatially disjoint but mutually linked, with winding numbers of 2 and 6, respectively. Between the two linked cases, the larger $m$ maintains the two closed components but increases the number of mutual turns, which is also reflected by the additional features in the orthogonal cross sections. The second viewing directions show that the apparent crossings arise from the mutual linking of two disjoint closed trajectories rather than from an overlap specific to a single projection. For $m = 3/4$ and $m = 5/4$, the reconstructed $|S_2|$ isosurfaces instead form a single closed tubular component in each case. The traced local maxima of $|S_2|$ form continuous 3D knot trajectories with winding numbers of 3 and 5, respectively. The larger $m$ produces a greater number of toroidal turns, while the alternative views make the continuous closure of each knot more apparent. The reconstructed experimental structures show local geometric deviations from the ideal results in Fig. 2, including variations in curvature, tube thickness, and transverse symmetry. These deviations may arise from residual optical aberrations, nonuniform spectral modulation, imperfect space-time matching between the two measured modes, finite sampling, interferometric drift, and the phase referencing between successive temporal slices. Despite these local deviations, the reconstructed trajectories preserve the designed closure relations and winding numbers. The agreement in closure and winding number is consistent with the experimental and theoretical trajectories belonging to the same topological class, despite their local geometric differences [5,9]. The reconstructed $S_2$ distributions retain the designed link-knot connectivity despite local geometric deviations from the ideal structures. Together with the Stokes channel separation described by Eqs. (3)-(6), these results support the polarization-cloaked representation of spatiotemporal links and knots.

## Discussion

We have established a polarization-cloaked representation of spatiotemporally localized optical links and knots by combining the dual phase winding of TLVs with the vectorial structure of the HOPS. In this construction, two orthogonally polarized components share a common toroidal envelope but carry opposite phase windings. Because the total intensity $S_0$ contains only the modal intensities, it keeps the spatiotemporal localization of the wavepacket without resolving the relative winding phase. By contrast, the phase sensitive Stokes parameters $S_1$ and $S_2$ contain the polarization interference terms and therefore retain the phase information that defines the link-knot connectivity. The same vectorial framework consequently supports different Stokes space connectivities while preserving the common total intensity envelope, and the 3D $S_2$ structures reconstructed from the experimentally retrieved TLV fields maintain the designed closure relations despite local geometric deviations. These results show that spatiotemporal localization and topological connectivity can be represented by different observables of the same electromagnetic wavepacket.

In scalar spatiotemporal links and knots, the topology is directly accessible through phase singularities or intensity related structures [1,4,19], whereas the present vectorial construction allows different link-knot connectivities to share the same total intensity envelope and become distinguishable through polarization-sensitive Stokes observables. Polarization therefore provides an additional degree of freedom for separating the spatiotemporal localization of a wavepacket from the representation and readout of its topology. This Stokes-resolved picture is also distinct from polarization singularity knots and vector mapping topologies such as optical skyrmions and hopfions [27,31,33]. Furthermore, the underlying mechanism is not restricted to the specific TLV family considered here: other vectorial spatiotemporal modes with matched envelopes and controllable relative phases may support similar polarization selective structures [44,45], while more general phase designs could extend the approach to higher-order links and knots [6]. Direct spatiotemporal Stokes measurements and studies of polarization crosstalk, noise, and propagation perturbations would further clarify the range and robustness of this mechanism. More broadly, these results extend spatiotemporal optical topology from a structure directly encoded in a scalar field to information that can also be represented and selectively accessed through the polarization degree of freedom.

## Acknowledgements

This work was supported by the National Natural Science Foundation of China (12534014 [Y.C.], 12192254 [Y.C.], W2441005 [Y.C.], 12434012 [Q.Z.], 62535013 [Q.Z.], 12604505 [X.L.], 12547149 [X.L.], 12622417 [C.L.], 12474336 [Q.C.]), the National Key Research and Development Program of China (2022YFA1404800 [Y.C.]), the Natural Science Foundation of Shandong Province (ZR2025ZD21 [Y.C.], ZR2023YQ006 [C.L.])). X.L. acknowledges support from the Young Talent of Lifting engineering for Science and Technology in Shandong (SDAST2026 QTA028) and the China Postdoctoral Science Foundation (2026M793716). S.A.P. acknowledges support from the Natural Sciences and Engineering Research Council of Canada (RGPIN-2025-04064).

## Author contributions

X.L and S.A.P conceived the idea and developed the theory. Y.Z and S.A.P verified the theoretical framework. Y.Z and N.Z performed the simulations and experiments. Y.Z, Z.Z, C.L, Q.C, H.T and X.L completed the data analysis and visualizations. Y.Z, S.A.P and X.L contributed to drafting and finalizing the manuscript. Q.Z, Y.C, S.A.P and X.L co-supervised the project. All authors participated in the discussion and review of the manuscript.

## Data availability

The data that support the findings of this study are available from the corresponding authors.

## Conflict of interest

The authors declare no competing interests.